\documentclass[aps,prl,twocolumn,showpacs,groupedaddress]{revtex4}

\usepackage[dvips]{graphicx}
\usepackage{amssymb,latexsym,pifont,psfrag}

\begin{document}

\title{Dynamical Janssen effect on  granular packings with moving walls}
\author{Yann Bertho, Fr\'ed\'erique Giorgiutti-Dauphin\'e and
Jean-Pierre Hulin} \affiliation{Laboratoire FAST, UMR 7608,
B\^at.~502, Universit\'e Paris-Sud, 91\,405 Orsay Cedex (France)}

\begin{abstract}
Apparent mass ($M_\mathrm{app}$) measurements at the bottom of
granular packings inside a vertical tube in relative motion are
reported. They demonstrate that Janssen's model is valid over a
broad range of velocities $v$. The variability of the measurements
is lower than for static packings and the theoretical exponential
increase of $M_\mathrm{app}$ with the height of the packing is
precisely followed (the corresponding characteristic screening
length is of the order of the tube diameter). The limiting
apparent mass at large heights is independent of $v$ and
significantly lower than the static value.
\end{abstract}

\pacs{45.70.-n, 81.05.Rm, 83.80.Fg}
\maketitle

\def\nl{\hfill\break}
\parindent= 15pt

Dense, dry vertical particle flows in channels are frequently
encountered in industrial processes such as the emptying of silos
or pneumatic transport \cite{Laouar98}. Particle fraction and
velocity variations in these flows depend largely on interaction
forces either between the particles or between the particles and
the channel walls \cite{Duran00b, Gennes99}. The present work
studies experimentally such forces in the case of a
granular packing globally at rest but in relative motion with
respect to an upwards moving tube. These force distributions have
been extensively studied by many authors for static or
quasi-static packing in tubes \cite{Janssen95, Bouchaud95,
Mounfield96, Mueth98, Pitman98, Vanel00, Ovarlez01} or other
configurations \cite{Pouliquen96, Edwards96b, Eloy97,
Rajchenbach01}. The pioneering work of Janssen demonstrated in
particular that the effective weight at the bottom end of
a container reaches exponentially a limit when the height of a
static packing increases: this is due to the shielding effect of
contact forces between grains redirecting the weight towards the
side walls. Vanel {\it et al.} \cite{Vanel99b} measured the
apparent weight of packed beads moving very slowly down a vertical
tube ($v$\,=\,20\,$\mu$m\,s$^{-1})$: these experiments indicate
that the weight increases during the displacement
before reaching a constant value but the influence of the height
of the packing was not investigated. In this work  we
demonstrate experimentally that Janssen's results can be
generalized to a grain packing moving with respect to solid walls
at velocities up to several cm\,s$^{-1}$.

\begin{figure}[h!]
\begin{psfrags}
\psfrag{vit}{$v$}
\includegraphics[width=8cm]{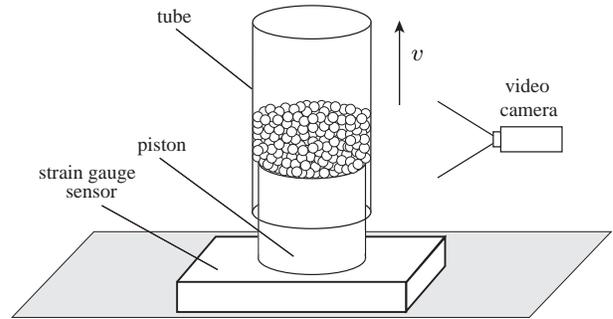}
\end{psfrags}
\caption{Sketch of the experiment.}
\label{dispositif}
\end{figure}

A packing of 2\,mm diameter glass beads (density
$\rho$\,=\,2.53\,10$^3$\,kg\,m$^{-3}$) is contained inside a
vertical glass tube of length 400\,mm and internal diameter
$D$\,=\,30\,mm. This tube can be moved up and down at a chosen
velocity $v$ ranging from 10\,$\mu$m\,s$^{-1}$ to
35\,mm\,s$^{-1}$. A cylindrical piston of height 80\,mm and
diameter 29\,mm is inserted in the bottom part of the tube: it is
carefully aligned with it and does not touch the walls during the
motion. The bottom end of the piston is screwed onto a strain
gauge sensor connected to a lock-in detector allowing one to
determine the force of the beads on the upper horizontal surface
of the piston (Fig.~\ref{dispositif}). This value is divided by
the gravitational acceleration $g$ to obtain an apparent mass
$M_\mathrm{app}$ determined with an uncertainty of $\pm$\,0.1\,g.
This apparent mass can be compared to the actual mass $M$ of the
grains measured by electronic scales prior to the experiment, with
a precision of $\pm$\,0.01\,g. In addition, qualitative
observations are realized using a video camera fitted with a macro
lens and connected to a video tape recorder. The experiments were
performed at a relative humidity $H$\,=\,(50\,$\pm$\,5)\%.
For each experimental run, the tube is first brought to its low
position and filled with a given mass of beads $M$, using a funnel
placed at the top of the tube. The relative motion between the
grains and the walls is then induced by moving the tube upwards
over a distance $\Delta z$\,$\simeq$\,70\,mm. In the initial and
final phases of the motion, the acceleration is equal to
0.04\,m\,s$^{-2}$ (or to 0.055\,m\,s$^{-2}$ for $v$ =
35\,mm\,s$^{-1}$) and the velocity $v$ is constant in between. At
the fastest velocity, the acceleration and deceleration distances
are of order 11\,mm and decrease to less than 10$^{-2}$\,mm for $v
<$\,1\,mm\,s$^{-1}$. One expects that, in the constant velocity
phase, the relative motion of the grains with respect to the walls
(and the force distribution) will be identical to those for a
steady downwards grain flow in a tube at rest. The two control
parameters of the experiment are the mass $M$ of the grains poured
into the tube and the tube velocity $v$. Compared to the flow of
grains inside a static tube, the relative motion of grains and of
the walls is the same but both the accelerations in the initial
and final phases and  the flow of air differ. Air flow should
however have a weak influence since pressure gradients in these
stationary compact grain flows are generally small.

\begin{figure}[h!]
\begin{psfrags}
\psfrag{ra}{({\it 1})} \psfrag{rb}{({\it 2})} \psfrag{rc}{({\it
3})}
\includegraphics[width=8cm]{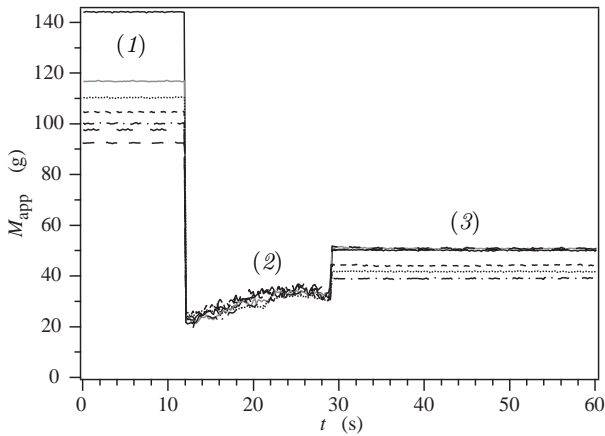}
\end{psfrags}
\caption{Variation of the apparent mass $M_\mathrm{app}$ of a
grain packing of total mass $M$\,=\,300\,g as a function of time:
({\it 1})~right after filling the tube, ({\it 2})~tube moving
upwards ($v$\,=\,4\,mm\,s$^{-1}$), ({\it 3})~tube stopped. Various
line patterns correspond to different experiments realized with
same control parameters.}
\label{Mappvst}
\end{figure}

Figure~\ref{Mappvst} displays variations of the apparent mass
$M_\mathrm{app}$ as a function of time for seven identical
experiments. The mass of the grains initially poured into the tube
is $M$\,=\,300\,g and corresponds to a height
$h$\,=\,(262\,$\pm$\,1)\,mm of the grain packing.
Values of the apparent mass $M_\mathrm{app}$ of the static packing
right after filling the tube (region ({\it 1}) in
Fig.~\ref{Mappvst}) are  dispersed: the  deviation is of the order
of $\pm$\,25\% of the mean value for large heights. Similar
observations are reported by  other authors
\cite{Duran98,Vanel99}: they are explained by the heterogeneous
stress network created when pouring the grains into the tube.
Arches of grains appear at random inside the packing and screen
the weight of other grains located above. These effects vary
strongly from one sample to another, resulting in large variations
of $M_\mathrm{app}$.

\begin{figure}[h!]
\includegraphics[width=8cm]{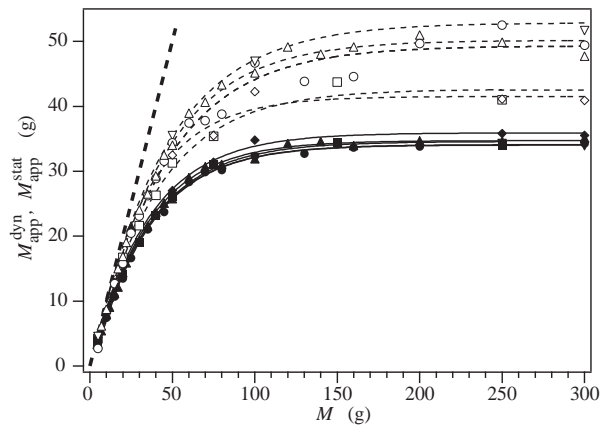}
\caption{Apparent mass as a function of total mass $M$ of grain
packing for different tube velocities: ($\lozenge,\blacklozenge$)
$v$\,=\,0.02\,mm\,s$^{-1}$, ($\square,\blacksquare$)
$v$\,=\,0.2\,mm\,s$^{-1}$, ($\circ,\bullet$)
$v$\,=\,2\,mm\,s$^{-1}$, ($\vartriangle,\blacktriangle$)
$v$\,=\,20\,mm\,s$^{-1}$, ($\triangledown,\blacktriangledown$)
$v$\,=\,34\,mm\,s$^{-1}$. Each line corresponds to an exponential
fit of the data points at each velocity. Dark symbols: limit value
$M_\mathrm{app}^\mathrm{dyn}$ at the end of region ({\it 2})
(moving tube) ($\Delta
M_\mathrm{app}^\mathrm{dyn}$\,=\,$\pm$\,1\,g). Open symbols:
apparent mass $M_\mathrm{app}^\mathrm{stat}$ in region ({\it 3})
(stopped tube) ($\Delta
M_\mathrm{app}^\mathrm{stat}$\,=\,$\pm$\,8\,g). Dashed thick line:
$M_\mathrm{app}$\,=\,$M$.} \label{Mappvsm}
\end{figure}

At the onset of the tube motion, the apparent mass decreases
abruptly and a random motion of the beads is observed. Then,
$M_\mathrm{app}$ increases slightly during the constant velocity
phase (region ({\it 2}) in Fig.~\ref{Mappvst}) and reaches a
limiting value referred to in the following as
$M_\mathrm{app}^\mathrm{dyn}$. A striking feature is the fact that
$M_\mathrm{app}^\mathrm{dyn}$ does not vary by more than
$\pm$\,5\% of its mean value from one sample to another.
Figure~\ref{Mappvst} also shows that the time dependence is the
same for all experiments within experimental uncertainty. When the
tube stops (region ({\it 3}) in Fig.~\ref{Mappvst}), the apparent
mass increases sharply to a constant value
$M_\mathrm{app}^\mathrm{stat}$ that is still much lower than in
phase ({\it 1}). The relative dispersion of these last static
values is again quite large ($\pm$\,15\%). Moreover, one observes
that the values of the apparent mass in region ({\it 1}) and
region ({\it 3}) are not correlated. These experiments were
repeated for different total masses $M$ of grains and different
tube velocities $v$. The respective experimental mean final values
of $M_\mathrm{app}$ after it has become constant in region ({\it
2}) and after the tube has stopped (region {\it 3}) are plotted in
Fig.~\ref{Mappvsm}.

The apparent mass $M_\mathrm{app}^\mathrm{dyn}$ at the end of the
moving phase (dark symbols), is almost independent of the velocity
for all total masses $M$ of grains in the tube. At low values of
$M$, one has $M_\mathrm{app}^\mathrm{dyn} \simeq M$ (there is no
screening effect). Then the curve levels off and
$M_\mathrm{app}^\mathrm{dyn}$ reaches a limit value
$M_\infty$\,=\,(34.5\,$\pm$\,1)\,g, independent of $v$.

After the tube motion has stopped (region {\it 3}), the apparent
mass increases by 20 to 60\% (open symbols) as already noted on
Fig.~\ref{Mappvst}. This increase is larger at higher velocities,
but the final value is always lower than in region ({\it 1}). The
global shape of the variation of $M_\mathrm{app}^\mathrm{stat}$
with $M$ is qualitatively similar to that obtained for
$M_\mathrm{app}^\mathrm{dyn}$, although the dispersion of the data
points is much higher.

Video recordings of the grain packing during the experiments
allowed one to estimate variations of the mean global particle
fraction $c$ in the sample: no  systematic motion of the upper
interface was detected and $c$ remained constant with
$c$\,=\,(64\,$\pm$\,0.5)\% in all experiments. These results
indicate that the motion of the walls reorganizes the internal
structure of the column: it allows to reach always a same
statistical dynamical equilibrium in spite of the variability of
the force distribution in the initial grain packing. This
reorganization results from very minute motions since no variation
of $c$ is detected. Bringing the walls to rest perturbs the stress
distribution and leaves the system ``frozen" in a state which may
vary significantly from one sample to another. As could be
expected, the perturbation is larger at high velocities.

The curves of Fig.~\ref{Mappvsm} follow {\it qualitatively} the
predictions of Janssen's model for static packing. We investigated
therefore whether the {\it quantitative} predictions were also
verified. The model assumes that the weight of the beads is
balanced by variations of the stress $\sigma _{zz}$ in the
vertical direction $z$ and by wall friction with:
\begin{equation}
{d\sigma _{zz} \over dz} + {1 \over \lambda} \sigma _{zz} =\rho
cg, \label{eq:equilibre}
\end{equation}
where $c$ is the local particle fraction of beads and
$\lambda$\,=\,$D / 4\mu K$ a characteristic length. The
coefficient $K$ characterizes the deflection of vertical stresses
into horizontal radial ones in the packing and $\mu$ is the
Coulomb friction coefficient. Integrating between the bottom
($z$\,=\,$z_0$) and the surface ($z$\,=\,0) of the packing where
$\sigma _{zz}(0) = 0$, gives the stress $\sigma _{zz} (z_0)$ on
the piston:
\begin{equation}
\sigma_{zz}(z_0) = \rho cg\lambda (1-e^{-z_0/\lambda}).
\label{eq:janssen}
\end{equation}
Since $M_\mathrm{app}$\,=\,$\sigma _{zz}(z_0)\pi D^2/4g$ and
$M$\,=\,$\rho c z_0\pi D^2 /4$, it follows that:
\begin{equation}
M_\mathrm{app} = M_{\infty} (1-e^{-M/M_{\infty}}),
\label{eq:janssenM}
\end{equation}
where $M_\infty$ is the predicted asymptotic value of the apparent
mass $M_\mathrm{app}$ when $M$ is increased:
\begin{equation}
M_\infty = \rho c\lambda \pi D^2 /4.
\label{eq:Minfini}
\end{equation}

\begin{figure}[h!]
\begin{psfrags}
\psfrag{vit}{$v$}
\includegraphics[width=8cm]{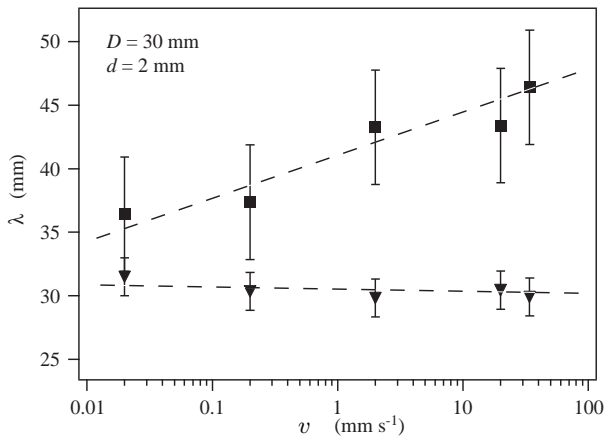}
\end{psfrags}
\caption{Janssen's length variations with the tube velocity:
($\blacktriangledown$) during the flow, ($\blacksquare$) after the
tube stops.} \label{Lvsv}
\end{figure}

The solid and dashed lines in Fig.~\ref{Mappvsm} represent
respectively the best fit with Eq.~(\ref{eq:janssenM}) of
experimental data points obtained during and after the motion of
the tube. The fits are excellent in the first case and fair in the
second one due to the larger dispersion of the data points. These
fits allow one to determine the asymptotic apparent mass
$M_{\infty}$ and the characteristic length $\lambda$ of Janssen's
model. Variations with the tube velocity $v$ of $\lambda$ values
corresponding to both sets of data are displayed in
Fig.~\ref{Lvsv}.
In the moving phase, $\lambda$ remains at the nearly constant
value $\lambda$\,=\,(30\,$\pm$\,2)\,mm. This value is close to the
tube diameter ($D$\,=\,30\,mm) and two or three times lower than
estimates in region ({\it 1}) from Eq.~(\ref{eq:Minfini}). After
the tube has been stopped (region {\it 3}), $\lambda$ is nearly
the same as in region ({\it 2}) at the lowest velocity but
increases by 30\% at the highest one. This confirms that
perturbations of the stress distribution induced by stopping the
tube are larger at higher velocities. A part of the arches and of
the contacts between grains probably disappear, which reduces the
value of the coefficient $K$ (and therefore increases $\lambda$).

\begin{figure}[h!]
\includegraphics[width=8cm]{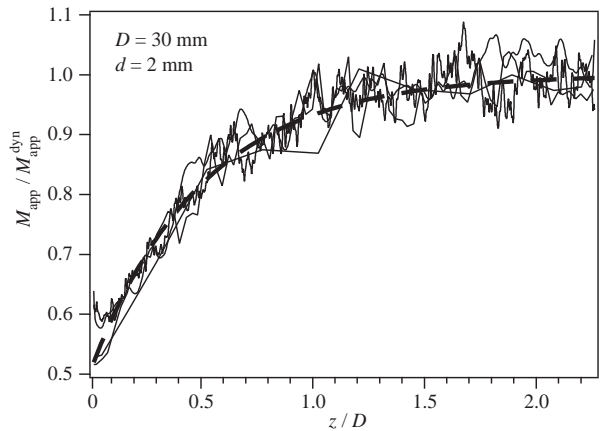}
\caption{Variation of the normalized apparent mass
$M_\mathrm{app}/M_\mathrm{app}^\mathrm{dyn}$ with the normalized
displacement $z/D$, at velocities 40\,$\mu$m\,s$^{-1}$\,$\leq
v\leq$\,34\,mm\,s$^{-1}$.} \label{Mappvsz}
\end{figure}

Another important issue is the dynamics of the grain rearrangement
while the tube is moving. This can be inferred from variations of
the apparent weight during the motion of the tube.
Figure~\ref{Mappvsz} displays superimposed variations of
$M_\mathrm{app}/M_\mathrm{app}^\mathrm{dyn}$ as a function of the
normalized displacement $z/D$ for several velocities ranging from
40\,$\mu m$\,s$^{-1}$ to 34\,mm\,s$^{-1}$. The total mass of
grains is equal to $M$\,=\,300\,g for which
$M_\mathrm{app}^\mathrm{dyn}$\,=\,$M_\infty$\,=\,34.5\,g.
A common increasing trend is followed and is well adjusted by an
exponential variation (dashed line):
\begin{equation}
{M_\mathrm{app}\over M_\mathrm{app}^\mathrm{dyn}} = 1-A \exp \left
(-{\alpha z\over D}\right ), \label{eq:fit}
\end{equation}
with $A$\,=\,0.48 and $\alpha$\,=\,2\,$\pm$\,0.2. The
characteristic relaxation length towards the limit value
$M_\mathrm{app}^\mathrm{dyn}$ in Fig.~\ref{Mappvsz} is equal to
$D$/2. The sharp drop of $M_\mathrm{app}$ right after the onset of
the motion implies that grains are initially dragged upwards
(a small transient motion is observed on video recordings);
then, the packing rearranges and more contact paths build-up
towards the sensor at the bottom.

The influence of the tube diameter $D$ has been investigated by
performing the experiments on a tube of larger diameter
$D$\,=\,40\,mm (beads of diameter $d$\,=\,3\,mm are used to keep
$D/d$ roughly constant).
\begin{figure}[h!]
\includegraphics[width=8cm]{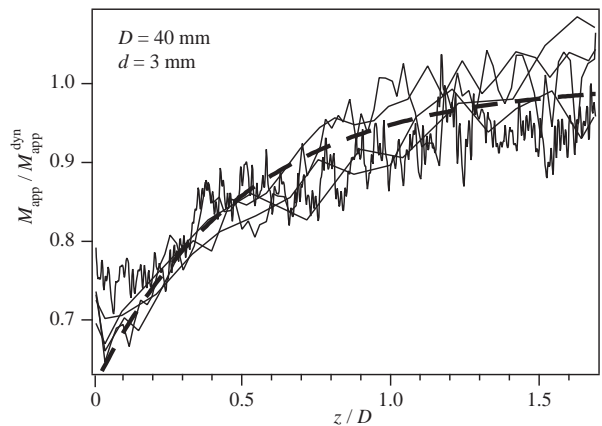}
\caption{Variation of the normalized apparent mass
$M_\mathrm{app}/M_\mathrm{app}^\mathrm{dyn}$ with the normalized
displacement $z/D$, at velocities 0.4\,mm\,s$^{-1}$\,$\leq
v\leq$\,34\,mm\,s$^{-1}$ ($M$\,=\,300\,g,
$M_\mathrm{app}^\mathrm{dyn}$\,=\,$M_\infty$\,=\,98\,g).}
\label{Mappvsz2}
\end{figure}
As for $D$\,=\,30\,mm, the variation of
$M_\mathrm{app}/M_\mathrm{app}^\mathrm{dyn}$ with $z/D$ is almost
independent on the tube velocity (Fig.~\ref{Mappvsz2}).
$M_\mathrm{app}$ increases also exponentially with the
displacement $z$, following Eq.~(\ref{eq:fit}) with the same value
of the parameter $\alpha$, as for $D$\,=\,30\,mm ($A$\,=\,0.39 and
$\alpha$\,=\,2\,$\pm$\,0.3). This indicates that the rearrangement
depends mostly on the normalized displacement $z/D$. The variation
of $M_\mathrm{app}$ with $M$ also verifies
Eq.~(\ref{eq:janssenM}): the corresponding shielding length is
independent of velocity with $\lambda  = (48 \pm 4)$ mm. As
expected, $\lambda$ increases with $D$; experiments for a broader
range  of values of $d$ and $D$ will be necessary to establish the
dependence of $\lambda$ on these parameters and on the ratio
$D/d$.

Preliminary experiments were performed at a lower relative
humidity ($H\lesssim$\,40\%). On the one hand, the variations of
$M_\mathrm{app}$ with the total mass $M$ remain in agreement with
Janssen's model and both the mean $M_\mathrm{app}$ values and
their relative dispersion are still much smaller than for the
initial static packing. These mean values are also of the same
order as before. On the other hand, $M_\mathrm{app}$ remains
constant in region ({\it 2}) instead of increasing towards a limit
value as previously. At the same time, grains remain at rest on
the outside of the packing while the tube moves. These results
imply that friction forces on the grains are smaller for
$H\,\simeq$\,40\%; they also indicate that even invisible
rearrangements of the packing during the wall motion may induce
large changes of the force distribution. Note that all
experimental data presented here were obtained after an
equilibrium of humidity between air and the grains had been
reached.

These results indicate that the grain packing reaches a dynamical
equilibrium  independent of the initial state and of the relative
velocity with respect to the walls: frequent transitions between
different force distributions in the packing are induced by the
tube motion leading to a well defined average value. These
distributions are  sensitive to small motions of the grains: these
do not induce measurable variations of the mean particle fraction
and are not even observable visually in some cases. The dynamical
equilibrium is reached exponentially with a characteristic length
proportional to the tube diameter: this may be due to the fact
that the reorganization of the grains starts at the walls and
propagates thereafter towards the center of the tube. Stopping the
flow increases $M_\mathrm{app}$, particularly at large velocities,
but it remains smaller than the initial value after filling. The
variability of $M_\mathrm{app}$ from one experiment to another is
much greater than during the tube motion since the system is
``frozen" in one state and the averaging effect of the variations
of the force distributions has disappeared.

To conclude, the present experiments demonstrate that Janssen's
model remains valid up to velocities of several centimeters per
second for granular packing inside a vertical tube in motion with
respect to the grains. The apparent mass $M_\mathrm{app}$ measured
at the bottom of the packing during the tube motion is much
smaller than for the initial static packing. The limiting value of
the apparent mass at the end of the moving phase
($M_\mathrm{app}^\mathrm{dyn}$) is also independent of the tube
velocity $v$ and much less variable from a sample to another than
for static packings (for a given total mass of the grains).
Globally, the result of this study suggests that force
distributions in dense granular flows may be described by
straightforward models over a much broader range of velocities
than usually expected.

We thank B. Perrin for many helpful discussions, H. Auradou and L.
Biver for their contribution to the experiments and G. Chauvin,
Ch. Saurine and R. Pidoux for the realization of the experimental
setup. We thank Ph. Gondret for a thoughtful reading of the
manuscript.

\bibliography{../../biblio/biblio}

\end{document}